\documentclass[journal,article,accept,moreauthors,pdftex,10pt,a4paper]{mdpi} 
\usepackage{amssymb,amsmath,bm}

\usepackage{epsfig}
\usepackage{bm}
\usepackage{color}
\usepackage{graphicx}

\firstpage{1} 
\articlenumber{x}
\doinum{10.3390/------}
\pubvolume{xx}
\pubyear{2017}
\copyrightyear{2017}
\externaleditor{Academic Editor: name}
\history{Received: date; Accepted: date; Published: date}

\Title{Hydrodynamics of a Granular Gas in a Heterogeneous Environment}

\Author{Francisco Vega Reyes $^{1}$ and Antonio Lasanta $^{2}$}

\AuthorNames{Francisco Vega Reyes and Antonio Lasanta}

\address{%
$^{1}$ \quad Departamento de F\'isica, Universidad de Extremadura, 06071 Badajoz, Spain 1; fvega@unex.es\\
$^{2}$ \quad G. Mill\'an Institute, Fluid Dynamics, Nanoscience and Industrial Mathematics
Department of Materials Science and Engineering and Chemical Engineering, Universidad Carlos III de Madrid, Legan\'es, Spain 2; alasanta@ing.uc3m.es}

\corres{Correspondence: alasanta@ing.uc3m.es}

\abstract{We analyze the transport properties of a low density ensemble of identical
macroscopic particles immersed in an active fluid. The particles are modeled as
inelastic hard spheres (granular gas). The non-homogeneous active fluid is modeled by
means of a non-uniform stochastic thermostat. The theoretical results are validated
with a numerical solution of the corresponding the kinetic equation (direct
simulation Monte Carlo method). We show that a steady flow in the system that is
accurately described by Navier-Stokes (NS) hydrodynamics, even for high
inelasticity. Surprisingly, we find that the deviations from NS hydrodynamics for
this flow are stronger as the inelasticity decreases.  The active fluid action is
modeled here with a non-uniform fluctuating volume force.  This is a relevant result
given that hydrodynamics of particles in complex environments, such as biological crowded environments, is still a question under intense debate.}


\begin{document}


\section{Introduction}

Biological Systems, and more specifically, systems with active particles, are in
general out of equilibrium, which implies that for these systems the results of equilibrium statistical
mechanics do not in general apply. The term \textit{active matter} refers to a
system with an ensemble of particles that are able to transform energy - internal or
environmental- into movement \cite{marchetti13}. A correct description of transport
phenomena in this type of systems can be achieved \textit{via} a specific kinetic
theory, that takes into account the peculiar energy processing of these particles, see for example \cite{bertin09,bertin06,chou12,ihle16} for the well known Vicsek-like models \cite{vicsek95}.

In recent years, the study of active particles in crowded environments
has become the subject of intense study \cite{bechinger16}. One of the
motivations is the large number of promising applications derived from
the use of active particles as micro-machines \cite{bechinger16}. In
addition, the behaviour of inert particles inside complex active environments remains an open question. For instance, there are experimental studies on inert Brownian particles trapped by a harmonic potential in a fluid composed by bacteria \cite{argun16}. Also, S\'andor et al. \cite{sandor17}  study active Brownian particles on a travelling-wave substrate. Other works study particles which can be trapped by holes \cite{sandor217,sandor317}. More realistic active systems such as biological tissues, share important analogies with granular matter, namely glass transition, cage effect, fluidization, solidification and the appearance of giant density fluctuations \cite{angelini11,malinverno17,garcia15,bi15}. There, active complex non-homogeneous forces appear. Also the non-Gaussian behaviour of diffusing  non-active particles in heterogeneous media has been studied recently for a spatially varying diffusion coefficient  \cite{malgaretti16}.

On the other hand, the transport properties of granular matter are of interest at a
fundamental level and also for applications, since granular matter is present in a
variety of industry and technology sectors \cite{D01,BP04,G03,dB11}. Granular
transport theories usually make a connection with traditional continuum mechanics
theories, due to fundamental similarities observed with classical fluids
\cite{D01,G03}. Granular particles loose a fraction of their kinetic energy
after collision and for this reason we need to
continuously excite the system, in order to sustain the dynamics. This excitation
usually comes from the system boundaries, which results in inhomogeneous flows that may
eventually become steady \cite{B53,D01,G03}.

We propose in this work a study that tries to model a system of macroscopic inert particles that are
immersed in an active fluid with inhomogeneous temperature. For this, we  model the system as a granular fluid \cite{H83} subject to a nonuniform stochastic volume force (in the form of a white noise \cite{PG12}). The inelastic cooling term, inherent in a granular fluid \cite{G03}, and
a stochastic force allow us in this work to model the energy sink and source terms that
are characteristic to active particle systems \cite{marchetti13,bechinger16}. More specifically, we focus on a very low density granular system (i.e., a
\textit{granular gas}) surrounded by a low density (inhomogeneous) fluid that can be considered as a coarse grained version of an active matter system. Since the system has very low density, particle velocity correlations can be neglected and thus the
the inelastic Boltzmann
equation applies \cite{G03,puglisi14}). Moreover, systems of solid particles
suspended in a low density interstitial fluid usually fall into the grain-grain collision dominated regime \cite{JN92}, as opposed to the interstitial fluid viscosity-dominated regime, that applies for most cases of interstitial liquids \cite{GLG99}.


In previous theoretical works, granular gas fluidization has been studied only in the simplistic case of a homogeneous interstitial fluid
\cite{WM96,CSH00,VPBTW06,VPV08,FAZ09,GSVP11}. And yet this
theoretical homogeneous state is not quite a realistic situation \cite{LBLG00}. Neat examples of inhomogeneous granular gas suspensions are the large sand plumes
that can be observed in the Earth's atmosphere for weeks \cite{Yea15}. A homogeneous energy source is obviously
not a realistic situation either in active fluids, at a biological level \cite{marchetti13}. For this reason we focus in this work on the more elaborate case of a non-uniform interstitial fluid.

However, the intent of this work is not experimental validation but to extract a more clear picture of the relevant physics of this kind of systems. 
As a result, we show there are granular gas flows immersed in an inhomogeneous active fluid hat can be accurately described with Navier-Stokes (NS)
hydrodynamics, even at high inelasticities. Furthermore, we
surprisingly found cases where NS hydrodynamics is a better
approximation for the granular gas than for a perfectly elastic gas subjected to the same boundary forcing conditions.
In particular, we will show that, once the granular gas is fluidized, the intensity
of the spatial gradients in response to external excitations (those coming from the
boundaries) can actually be weaker for granular gases than for elastic gases.

In a previous experimental work \cite{OLDLD04}, local
balance between the total energy input (fluctuating force plus boundary heating) and
the energy sink was found. We use now this kind of balance condition. As we will show below, this local balance results, in the specific
set-up considered in the present work, in a steady flow with uniform heat flux
throughout the system. This balance condition is in fact analogous to the one occurring in well known non-Newtonian granular flows like the uniform shear
flow \cite{C90,VSG10}.

The structure of the paper is as follows: in Section 2 we present a description of
the system and the corresponding steady state equations. In section 3 we describe the
flows where there is a balance between inelastic cooling coming from particle
collisions and volume energy input (and with no shear). Section 4 is devoted to a
brief discussion on the steady flows resulting from adding a weak shear Section 5
presents a discussion on the transport coefficients and rheology properties of the
studied flows and the paper conclusions are drawn in Section
6. The simulation method and transport coefficient equations  are
discussed in the Appendices \ref{comp} and \ref{coef} respectively.
 
\section{System and Steady base state equations}
\label{sys}

\subsection{Description of the system}

Our system consists of a set of identical inert smooth hard spheres (or disks)
immersed in an active fluid. The system is limited by two infinite parallel hard walls, located at planes $y=\pm L/2$ respectively. The walls act like two distinct kinetic energy sources, characterized with temperatures $T_\pm$. They may have relative movement (wall velocities $U_\pm$ respectively), eventually inducing the particles to continuously flow along the channel between them.

Collisions between the inert particles do not preserve
energy (i.e., particle collisions are \textit{inelastic}). The particles have a diameter (radius)
that we denote as $\sigma$ and their mass is $m$. For inelastic smooth hard
spheres/disks, inelasticity may be characterized accurately, in a range of
experimental situations, by a constant coefficient of normal restitution $\alpha$
\cite{FLCA94,GBM15}. This coefficient ranges from 1 for purely elastic collisions to
0 for purely inelastic collisions. In addition, we model the interstitial
active fluid injection of thermal energy onto the granular gas particles as a
stochastic force $\mathbf{F}^\mathrm{st}$. The equation of motion for a particle $i$ can be written

\begin{equation}\label{eqmov}
m {\bf \dot{v}}_i= {\bf F}_i^{\text{st}}( {\bf r},t)+ {\bf F}_i^{\text{coll}},
\end{equation} where $ {\bf F}_i^{\text{coll}}$ is the force due to inelastic
collisions,  ${\bf F}_i^{\text{st}}({\bf r},t)$ is the force exerted by the
heterogeneous medium and $m$ the mass of particles (set to 1 for simplicity).

We can model this interaction by means of a fluctuating volume force \cite{K92}, that we denote as $\mathbf{F}^\mathrm{st}({\bf r})$ and fulfills the conditions \cite{MTC07,GSVP11,K92,OLDLD04}

\begin{eqnarray}
& &\langle {\bf F}_i^{\text{st}}({\bf r},t) \rangle ={\bf 0},\: \nonumber \\ 
& &\langle {\bf F}_i^{\text{st}}({\bf r},t) {\bf F}_j^{\text{st}}({\bf r},t') \rangle =\mathbf{1}m^2 \xi({\bf r})^2 \delta_{ij}\delta(t-t'),
\label{Fst}
\end{eqnarray} being $\xi^2 ({\bf r})$ the fluctuating force intensity
\cite{MTC07,GSVP11} (that has dimensions of velocity squared times time),
$\mathbf{1}$ the unit matrix in $d$ dimensional space, $\delta_{ij}$ is the Kronecker
delta function and  $\langle A\rangle$ indicates average of magnitude $A$ over a
sufficiently long time interval (long compared with the characteristic microscopic
time scale).  It has been shown in the case of the large mass limit and homogeous energy injection, that equation \ref{eqmov} for a particle can be written as a Langevin equation corresponding to a granular brownian particle \cite{Sarracino10}.

On the other hand, collisions between two grains follow the inelastic smooth hard
sphere collisional model
\begin{eqnarray}
& & {\bf v}'_1={\bf v}_1-\frac{1+\alpha}{2} ( \pmb{\sigma } \cdot {\bf g}) \pmb{\sigma } \\
& &  {\bf v}'_2={\bf v}_2-\frac{1+\alpha}{2} (\pmb{\sigma } \cdot  {\bf g}) \pmb{\sigma }
\end{eqnarray}
with $\pmb{\sigma }=( {\bf r}_1-{\bf r}_2)/|{\bf r}_1-{\bf r}_2|$ and $ {\bf g}={\bf v}_1-{\bf v}_2$. The primes denote pre-collisional velocities.

Notice that in \eqref{Fst} we have $\xi=\xi({\bf r})$ (a space-dependent noise intensity).  The noisy term appears in the inelastic Boltzmann equation as a Fokker-Planck-like operator \cite{GCV13,MPRV08}

\begin{equation}
\left(\frac{\partial}{\partial t}
+\mathbf{v}\cdot\nabla-\frac{1}{2}\xi({\bf r})^2\frac{\partial^2}{\partial v^2}\right)f(\mathbf{r},\mathbf{v};t)=J[\mathbf{v}|f,f], 
\label{BE}
\end{equation} 
where $J$ is the collisional integral for inelastic hard spheres and whose expression may be found elsewhere \cite{G03,BDKS98}, $\mathbf{v}$ is particle velocity, and $v$ its modulus.

Taking into account the system geometries, we consider only $\xi({\bf r})=\xi(y)$,
that is the simplest space-dependence our geometry configuration allows for an
inhomogeneous active fluid. The steady base states may be deduced from the reduction of the kinetic equation \eqref{BE} 

 \begin{equation}
v_y\frac{\partial f}{\partial y}-\frac{1}{2}\xi(y)^2\frac{\partial^2f}{\partial v^2}=J[f,f].
\label{rBE}
\end{equation} 

We have numerically solved the kinetic equation \eqref{rBE} by means of the Direct Simulation Monte Carlo method \cite{BP04,B94}, applying in each case the corresponding boundary conditions and heterogeneous fluctuating force properties. For more details on the simulation algorithm, see the corresponding section in the Appendix \ref{comp}.

\subsection{Steady base state equations}
Multiplying by velocity momenta the kinetic equation \eqref{rBE} and performing velocity integrals, we may easily obtain from \eqref{rBE} the mass, momentum and energy balance equations \cite{VU09,GCV13}

\begin{eqnarray}
& & \frac{\partial P_{yy}}{\partial y}=0,\quad 0=-\frac{1}{mn}\frac{\partial P_{xy}}{\partial y} \label{balPy}
\\
& & T\zeta (\alpha)-m\xi^2(y)=-\frac{2}{dn}\left(P_{xy}\frac{\partial u_x}{\partial y}+\frac{\partial q_y}{\partial y}\right), \label{balTy}
\end{eqnarray} 
where $\mathbf{u}$ is the flow velocity field  (with $u_y$ its y-component), $n$ is the particle density, and $P$ and $q$ are momentum flux tensor (also called stress tensor) and heat flux vector respectively, and $\zeta(\alpha)$ is the cooling rate, a magnitude that emerges from the energy balance due to inelasticity in the collisions \cite{G03,BP04}. Equations \eqref{balPy} and \eqref{balTy} contain no approximations; i. e. they are exact for this type of geometry, even if the steady state were not hydrodynamic. Notice also that the stress tensor element $P_{yy}$ is homogeneous and is not a function of time. This condition breaks if the flow is not laminar, but we will only consider laminar flows in this work. 

We take now into account the characteristics of NS hydrodynamics in \eqref{balPy} and \eqref{balTy}. In \eqref{balPy} we use that all diagonal elements of the stress tensor are equal and we also use that the horizontal heat flux $q_x$ is null (they both raise from fluxes terms that are of second order in the gradients \cite{B35}) and thus, from the first equation in \eqref{balPy}, we obtain that $P_{yy}=p=\mathrm{constant}$. We also use the fluxes expressions at NS order \cite{BDKS98}: $P_{xy}=\eta(\alpha)\partial u_{ x}/\partial y$, $q_y=-\lambda\partial T/\partial y-\mu\partial n/\partial y $, where $\eta$ is the viscosity and $\lambda, \mu$ are the heat flux transport coefficients. With this, the steady state forms of \eqref{balPy} and \eqref{balTy} are
\begin{eqnarray}
& &p =\mathrm{constant},\quad  P_{xy}=-\eta (\alpha)\frac{\partial u_x}{\partial y}=\mathrm{constant} \label{balNSPy}
\\
 & &T\zeta (\alpha)-m\xi^2(y)=  \nonumber\\
& &\frac{2}{dn}\left(\eta (\alpha) \left(\frac{\partial u_x}{\partial y}\right)^2+\kappa (\alpha)\frac{\partial T}{\partial y}+\mu (\alpha) \frac{\partial n}{\partial y}\right). \label{balNSTy}
\end{eqnarray}

Next, we take into account that the NS transport coefficients and cooling rate scale in the following forms with temperature \cite{BDKS98}: $\eta (\alpha)=\eta^*(\alpha)\eta_0'T^{1/2}$, $\kappa (\alpha)=\kappa^*(\alpha)\kappa_0'T^{1/2}$, $\mu=\mu^*(\alpha)\kappa_0' T^{3/2}/n$, $\zeta=\zeta^*(\alpha)p/(\eta_0'T^{1/2})$. Here, the coefficients $\eta_0'$, $\kappa_0'$ are related to the elastic gas  \cite{CC70} NS viscosity $\eta_0=\eta_0'T^{1/2}$ and thermal conductivity $\kappa_0=\kappa_0'T^{1/2}$, and thus
\begin{eqnarray}
& &\eta_0'=\frac{d+2}{8}\Gamma (d/2)\pi^{-(d-1)/2}m^{1/2}\sigma^{-(d-1)}/8, \\ 
& & \kappa_0'=\frac{d(d+2)^2}{16(d-1)}\Gamma (d/2)\pi^{-(d-1)/2}m^{-1/2}\sigma^{-(d-1)}.
\end{eqnarray}where $\Gamma$ is the gamma function \cite{AS65} and $d=2$ for disks and $d=3$ for spheres. Thus, the steady state balance equations \eqref{balNSPy}, \eqref{balNSTy} yield

\begin{eqnarray}
& &-\eta(\alpha)^*\eta_0'\sqrt{\frac{T}{T_r}}\frac{\partial u_x}{\partial y}=\mathrm{constant}, \label{prev_profu} \\
& & \sqrt{\frac{T}{T_r}}\frac{\partial}{\partial y}\left(\sqrt{\frac{T}{T_r}}\frac{\partial T}{\partial y} \right) =  \nonumber \\
& & \frac{d}{2}\frac{\zeta_0^*}{\beta_0^*}\frac{p^2}{T_r}-\frac{d}{2}\frac{m\xi^2(y) nT^{1/2}}{\beta_0^*T_r}-\frac{\eta_0^*}{\beta_0^*}\frac{T}{T_r}\left(\frac{\partial u_x}{\partial y}\right)^2,  \label{prev_profT}
\end{eqnarray} where
$\beta^*(\alpha)\equiv\kappa^*(\alpha)-\mu^*(\alpha)$, and we call it
\textit{effective thermal conductivity} and $T_r$ is the reference
unit of temperature, that we choose as $T_r=T_-$ (temperature at the
lower, and colder, wall). Here, $\zeta^*(\alpha)$ is the dimensionless
cooling rate \cite{BP04,VSG13} and the expressions of the reduced
transport coefficients viscosity $\eta^*(\alpha)$, thermal
conductivity $\kappa^*(\alpha)$, and $\mu^*(\alpha)$  that we used are
the ones obtained by Garz\'o and Montanero \cite{GM02} for a granular
gas heated by a white noise. Their expressions are displayed in the
transport coefficients section in the Appendix \ref{coef}.

We use a scaled space variable $l$ that allows us to find an analytical solution for the flow velocity \eqref{prev_profu} and the temperature \eqref{prev_profT} profiles. This variable is defined by the relation $\sqrt{T/T_r}\partial /\partial y=\partial /\partial l$. When expressed in terms of this scaled variable, equations \eqref{balNSPy}-\eqref{balNSTy} read 

\begin{eqnarray}
& &\frac{\partial\hat u_{y}}{\partial\hat l}=a,
\label{Pij}\\
& &\frac{\partial^2 \hat T}{\partial\hat l^2}=-\gamma(\alpha, l),
\label{difT}
\end{eqnarray} where $a$ and $-\gamma$ are dimensionless magnitudes in the system and we call them \textit{local shear rate} and \textit{temperature curvature}, respectively. 
In \eqref{Pij}, \eqref{difT} and henceforth we use dimensionless variables indicated with a hat, where we have chosen as set of units: $T_r$ for temperature (already defined), $m$ for mass, $p$ for pressure (the steady state hydrostatic pressure, that as we said is a global constant for the system). As a unit for length we use 

\begin{equation}
\lambda_r=\sqrt{2}(2+d)\Gamma (d/2)\pi^{-(d-1)/2}(\sqrt{2}n_r\sigma^{(d-1)})^{-1},
\end{equation} that is the reference mean free path \cite{CC70} and for time we use

\begin{equation}
  \nu_r^{-1}=\lambda_r(\sqrt{m/T_r})
  \label{nur}
\end{equation} (inverse of reference collision frequency). These two definitions imply that the unit for velocity is $\sqrt{T_r/m}$. We indicate the spatial coordinate $y$ in dimensionless form as $\hat y$, and analogously for all dimensionless magnitudes. 

The reader may infer the physical meaning of $a$ and $\gamma$ with the aid of \eqref{balPy}, \eqref{balTy}: $a$ (reduced shear rate) measures how rapidly varies the flow velocity field, in units of the mean free path; $|-\gamma|^{1/2}$ measures the degree of imbalance between energy volume sinks and sources per unit time (as we explain above in Eq. \eqref{nur}, this unit being the inverse of the collision frequency). The differential equation \eqref{difT} has been observed for the granular temperature in similar experimental configurations \cite{LBLG00}.

With the presence of the stochastic volume force in the energy balance equation \eqref{balTy}, it is straightforward to deduce from \eqref{prev_profT}- \eqref{difT} and the definitions above that $-\gamma (\alpha, l)$ has the form

\begin{eqnarray}
  -\gamma(\alpha)\equiv & &\frac{2(d-1)}{d(d+2)\beta^*(\alpha)}\times \label{ebal} \\
& &\left(\frac{d}{2}(\zeta^*(\alpha) -\hat\xi^2(\hat l)\hat T(\hat l)^{-1/2})-\eta^*(\alpha)a^2\right),  \nonumber 
\end{eqnarray} that, as it can be noticed, depends in general on position through $\hat l$.

\section{Steady Base States with energy balance and no shear}
\label{reference}

We now want to look up for steady states in a granular gas fluidized by an heterogeneous source that may be accurately described by hydrodynamics at NS order. We will denote them as 'reference states'. As in rarefied gases, a close to unity Knudsen number ($\mathrm{Kn}$) signals hydrodynamics failure. Moreover, hydrodynamics at first order in the gradients (NS equations) are usually only valid if $\mathrm{Kn}$ takes low values \cite{G03,B94}. The Knudsen number $\mathrm{Kn}$ is defined as a representative microscopic time or length scale over a macroscopic time or length scale. However, the choice of reference scales is not unique and picking an appropriate Knudsen number for each specific flow problem is a subtle question \cite{BCC95}. 

It was shown in a previous work \cite{VU09} that Couette-Fourier flows (i.e., those
occurring in our system, under the appropriate circumstances) are characterized by
three representative and independent microscopic vs. macroscopic relative scales (or
Knudsen numbers) that happen to be constant throughout the system: $a$,
$|\gamma|^{1/2}$ (look at eqs. \eqref{Pij} and \eqref{difT}) and $\Delta T/L$,  with
$\Delta T\equiv T_+-T_-$, where $\Delta T/L$ measures the size of the temperature
gradient imposed by the boundaries, this gradient being referred to mean free path
units. These numbers arise from the specific form of the relevant differential
equations for steady states, and the boundary geometry. Since the geometry of the
differential equations \eqref{Pij}, \eqref{difT} and boundary conditions are the same
in this work, we may safely use the same set of reference Knudsen numbers (see
previous work \cite{VU09} for more details on this issue). Furthermore, another
previous analysis \cite{VSG13} for granular gases without an interstitial fluid
showed that $\Delta T/L$ has a much weaker effect on the emergence non-Newtonian
behavior.

Thus, let us analyze the class of solutions for steady flows where
this $\Delta T/L$ is the only active source of spatial gradients;
i.e., $a=0, \gamma=0$ and 
$\Delta T/L\lesssim 1$

\begin{eqnarray}
& &\frac{\partial\hat u_{y}}{\partial\hat l}=0,
\label{profu}\\
& &\frac{\partial^2 \hat T}{\partial\hat l^2}=0.
\label{profT}
\end{eqnarray} 

 Moreover, in the $\alpha=1$ limit this state corresponds to the classic Fourier flow of a molecular gas. In order to obtain $\gamma=0$ (with $a=0$), and taking into account \eqref{ebal} we find 

\begin{equation}
  \label{profnoise}
  \hat\xi(\hat l)^2\hat T(\hat l)^{-1/2}= \zeta^*(\alpha) \quad \Rightarrow \quad \hat\xi(\hat l)^2= \zeta^*(\alpha)\hat T(\hat l)^{1/2}.
\end{equation} 
Equation \eqref{profu} implies that flow velocity $\hat u$ is constant respect to $\hat l$, which, from the definition of $l$ leads to $\hat u(\hat y)$ also constant (and to all physical effects, flow velocity may be considered as null). Moreover, equation \eqref{profT} implies that temperature $\hat T$ is linear in $\hat l$, which (since $\sqrt{T/T_r}\partial/\partial y=\partial /\partial l$) yields $\hat y\propto \hat l^{3/2}$ and thus $\hat T(\hat y)\propto\hat y^{2/3}$; i.e., the stochastic force generates a Fourier-like temperature profile \cite{VSG10}. At a physical level, this makes sense if we consider that the granular gas is fluidized by a low viscosity interstitial fluid that is heated from the boundaries. It is easy to deduce from \eqref{balTy}, \eqref{balNSTy} and \eqref{ebal} that $\gamma=0$ implies that inelastic cooling is balanced by volume energy input and thus, that heat flux is uniform throughout the system. 

Therefore our base steady states are (not previously reported) granular flows with uniform heat flux. Summarizing, these NS steady states have the following properties: i) $P_{xy}=\eta(\alpha)\partial u_{ x}/\partial y$, $q_y=-\lambda\partial T/\partial y-\mu\partial n/\partial y $, ii) constant hydrostatic pressure $\hat p$, iii) null normal stress differences ($\hat P_{ii}=\hat P_{jj}=p$) and null $\hat q_x$ \cite{B35}, iv) constant $\hat q_y$, and v) linear $\hat T(\hat l)$ profiles, or equivalently, $\hat T(\hat y)\propto \hat y^{3/2}$ \cite{VSG10}. Properties i) to iii) are fulfilled by all Couette-Fourier steady flows at NS order \cite{VU09} and conditions iv) and v) are fulfilled by all Couette-Fourier steady flows with uniform heat flux \cite{VSG10}. 

We need to remark that the relation in \eqref{profnoise} between fluctuating force
intensity and temperature is not a hypothesis but a consequence of the experimental
fact that there is local energy balance. On the other hand, there is an essential
difference of this new flow respect to previously studied granular flows with uniform
heat flux (the simple shear flow included \cite{C89,C90}).



\begin{figure}
\centering
\includegraphics[height=6.1cm]{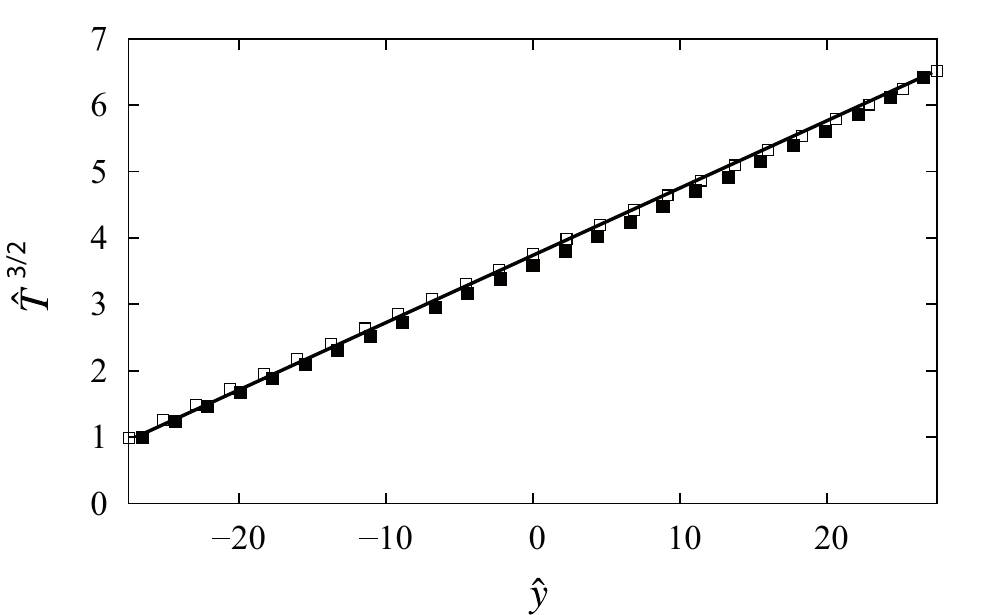}
\caption{DSMC simulation data ($d=3$, spheres) for temperature profile for two different values of inelasticity: $\alpha=0.9$ (solid symbols) and $\alpha=0.7$ (open symbols) and with the same boundary condition for wall temperature difference ($T_+/T_-=5$, $\Delta L=15\overline\lambda$) with $\overline\lambda=\sqrt{2}\pi\overline n\sigma^{d-1}$, $\overline n$ being the average density in the system. There is excellent agreement with the theoretical prediction of $\hat T(\hat y)\propto y^{2/3}$ (solid line stands for the NS theoretical profile), for both $\alpha$ values.} 
\label{figT}
\end{figure}

Very surprisingly, we can show that contrary to other steady granular flows, the kind
of state defined by \eqref{profnoise} can be accurately described in the context of Navier-Stokes equations. In order to prove this feature, we have simulated
the system described by equations \eqref{profu}, \eqref{profT} and \eqref{profnoise} by
means of the DSMC method (more details on simulations are given in the Appendix \ref{comp}). We have checked that the temperature profiles obtained under steady state in effect follow the behaviour $\hat T(\hat y)\propto y^{2/3}$, inherent to gas steady flows with uniform heat flux \cite{VSG10}. Results can be seen in Figure \ref{figT}, for which the agreement with the NS theory prediction is excellent, independently of the inelasticity value.

\section{Weakly sheared steady states}
\label{shear}

We discuss here weakly sheared states, in order to capture the gradual the eventual appearance of non-Newtonian behavior. Thus, starting out of the reference base states in the previous subsection; i.e., still with condition \eqref{profnoise} being fulfilled, we analyze now sheared states ($a\neq 0$), for which we have from \eqref{ebal}, 

\begin{equation}
-\gamma=-\frac{(d-1)}{(d+2)\beta^*(\alpha)}\eta^*(\alpha)a^2\neq 0.
\label{GRshear}
\end{equation} For this type of flow we have also performed DSMC
simulations. Relation \eqref{GRshear} helps us understand that an increase in shear
rate $a$ (one of the relevant Knudsen numbers) implies an increase in $\gamma$ (one
of the other Knudsen numbers). Thus, for increasing shear we should gradually depart
from NS hydrodynamics as both $a$ and $|\gamma|^{1/2}$ increase. However, relation
\eqref{GRshear} is valid only as long as NS hydrodynamics applies. For non-Newtonian
regime, the relation between the temperature curvature $\gamma$, local shear rate $a$
and coefficient of restitution $\alpha$ would be more intricate and in general it
must be solved numerically \cite{VSG13}. A non-linear theory is beyond the scope of
the present work, but, fortunately, we may use results from DSMC to analyze
deviations from NS hydrodynamics.

The results we obtained are quite surprising (see Table 1). In effect, we can see that similar values of local shear rate $a$ tend to yield greater values of the other Knudsen number $|\gamma|$ for more elastic gases than for more inelastic gases. This means that, contrary to what would be expected, the Knudsen numbers in this type of flow tend globally to be smaller for higher inelasticities, which could indicate that departures from NS hydrodynamics are more important, for similar shear rates, for more elastic gases. But we will confirm this point in the following section \ref{transport}, by measuring from DSMC simulations the set of transport coefficients and rheological properties and comparing with NS theory predictions.

\begin{table}
\centering
\caption{\label{tabla} DSMC measurements of Knudsen numbers $a$ and $|\gamma|^{-1/2}$, for steady states described by \eqref{profu}, \eqref{profT}, \eqref{profnoise}, with $T_+/T_-=5$, $\Delta L=15\overline\lambda$ (i.e., $\Delta T /\Delta L= (4/15) ((T_+-T_-)/\overline\lambda)$,  and $\Delta U$ defined as $\Delta U\equiv (U_+-U_-) /\sqrt{T_-/m}$.}
\begin{tabular}{lllll}
 $ $ & $\Delta U=5 $ & $\Delta U=5$& $ \Delta U=3$ & $\Delta U=3$   \\ 
$\alpha$ & $a$ & $|\gamma|^{-1/2}$& $a$ & $|\gamma|^{-1/2}$   \\

  0.99 & 0.173 & 0.058 & 0.115 & 0.036 \\
  0.9 & 0.191 & 0.034 & 0.120 & 0.018 \\
  0.8 & 0.193 & 0.027 & 0.119 & 0.017 \\
  0.7 & 0.193 & 0.024 & 0.119 & $\lesssim 1\times 10^{-2}$ \\
  0.6 & 0.188 &  $\lesssim 1\times 10^{-2}$ & 0.115 & $\lesssim 1\times 10^{-2}$ \\
  0.5 & 0.187 &  $\lesssim 1\times 10^{-2}$  & 0.113 &  $\lesssim 1\times 10^{-2}$ \\
\end{tabular}
\end{table}

We also performed an additional series with varying $\Delta T/\Delta L$ (not shown in figures) and checked that the measured transport coefficients do not depend on its value, analogously to the result in a previous work on granular flows with uniform heat flux \cite{VSG10}.


\section{Transport coefficients and rheology}
\label{transport}

Results in the preceding section suggest that: a) our steady states are well described by NS hydrodynamics, even for strong inelasticities, as long as shearing from the boundaries is not too strong; b) deviations from NS hydrodynamics are stronger for more elastic gases. Both observations would be contrary to what occurs in steady granular flows (like the simple shear flow \cite{C90}, just to put one example).

Let us analyze in more detail to what extent the reference (no-shear) steady states described in section \ref{reference} and the sheared states described in section \ref{shear} are strictly NS flows or not, by analyzing the relevant transport properties. This analysis will be done as a function of the degree of inelasticity in grains collisions and intensity of shearing. In order to analyze non-Newtonian behavior, we need to define the cross thermal $\phi^*$ conductivity transport coefficient $\phi^*$ and the reduced normal stresses, defined by 

\begin{equation}
  \label{rheol}
  q_x=\phi^*\lambda_0\partial T/\partial y, \quad \theta_i=P_{ii}/p,
\end{equation} where $\lambda_0$ is the thermal conductivity for a gas with elastic particle collisions \cite{CC70}. For hydrodynamics at NS order, $q_x,\phi^*=0$ and  $\theta_i=P_{ii}/p=1$. The degree of divergence from these numerical values quantifies eventual departures from NS hydrodynamics. The appearance of  $q_x\neq 0$ and $\theta_i\neq 1$ occurs with the emergence in the fluxes of terms of second order in the gradients (Burnett order). For our geometry, the only remaining second order terms are of form $(\partial T/\partial y)(\partial u_x/\partial y)$, $\partial^2u_x/\partial y^2$ in the heat flux and of the form $\partial^2 T/\partial y^2$, $(\partial T/\partial y)^2$, $(\partial u_x/\partial y)^2$ in the stress tensor (moment flux) respectively (the rest of the gradient second order terms in the fluxes are null for the geometry of our system) \cite{B35,CC70}.



\begin{figure}
\centering
\includegraphics[height=8.25cm]{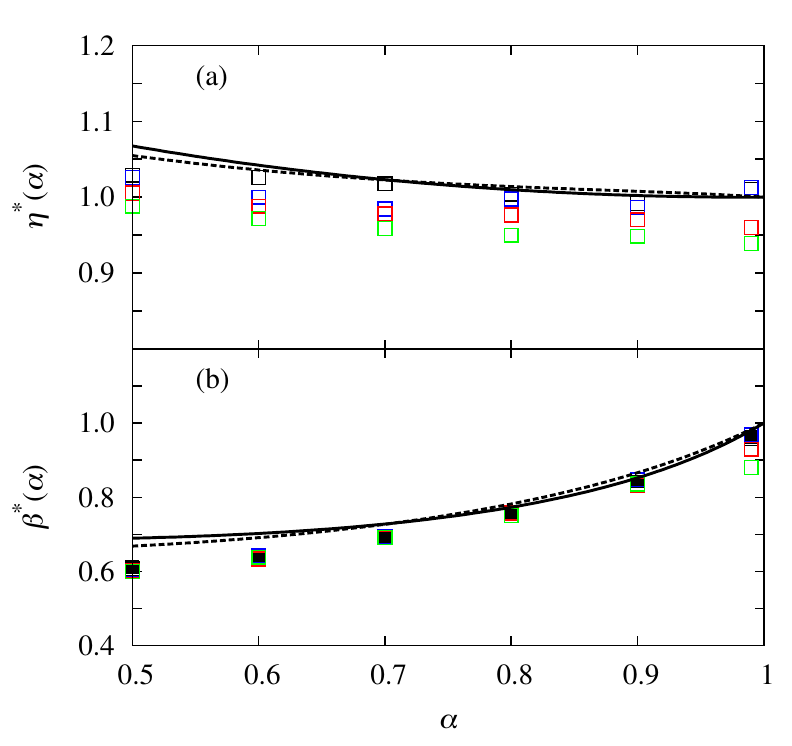}
\caption{(a) Viscosity vs. coefficient of restitution $\alpha$. (b) Effective thermal conductivity vs. $\alpha$. In both panels, lines represent the results for the theoretical NS coefficient (solid for standard Sonine approximation and dashed for improved Sonine approximation) whereas the symbols stand for DSMC simulations. Several DSMC series have been represented, with increasing relative wall velocity. Solid symbols represent the no-shear case and open symbols are the sheared states for: $\Delta U= 0.25$ with $a\simeq 0.019$ (black), $\Delta U=1$ with $a\simeq 0.04$ (blue) , $\Delta U=3$ with $a\simeq 0.10$ (red), $\Delta U=5$ with $a\simeq 0.18$ (green).} 
\label{figcoefs}
\end{figure}

In Fig. \ref{figcoefs} we present the theory-simulation comparison for reduced
viscosity $\eta^*(\alpha)$ and effective thermal conductivity $\beta^*(\alpha)$, for
different shearing levels. For sufficiently low shearing, the agreement between
theory and simulation is very good for the range of inelasticities represented
here. In the limit of no shear, the agreement for the thermal conductivity is
excellent from $\alpha=1$ down to $\alpha=0.7$. The viscosity, however, cannot be
measured on the reference (null shear) steady states. Of course, this limitation is not intrinsic of the granular gas since the same happens for classic fluids. Nevertheless, a relatively good viscosity measurement can be obtained in the series with lowest shear used in the simulations ($a\simeq0.019$), for which the agreement at high inelasticities between theory and simulation is actually slightly better than for the thermal conductivity measured at $a=0$. For both coefficients, and at high inelasticities ($\alpha\simeq 0.5$), the agreement is improved when the zeroth order distribution function in the Chapman-Enskog method \cite{CC70} is not the Maxwellian but an approximation to the solution of the homogeneous state for the heated granular gas (slightly different from the Maxwellian \cite{GSM07}). This approximation improve (represented with dashed lines in Fig. \ref{figcoefs}) consists in taking into account the first term in an expansion in Sonine polynomials around the Maxwellian (for more reference on the Sonine polynomials expansion method in kinetic theory, see for instance \cite{B35,CC70,T73}). 


\begin{figure}
\centering
\includegraphics[height=8.25cm]{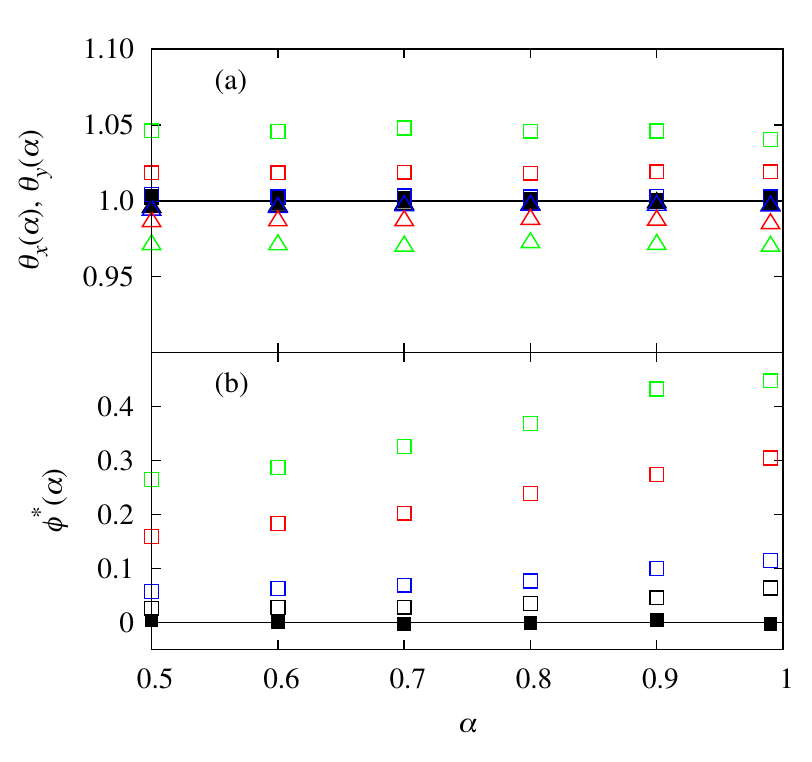}
\caption{(a) DSMC data for reduced normal stresses $\theta_x\equiv P_{xx}/p, \theta_y\equiv P_{yy}/p$ vs. coefficient of restitution $\alpha$. Square symbols stand for $\theta_x$ and triangles stand for $\theta_y$. (b) DSMC data for thermal cross conductivity vs.  $\alpha$. In both panels, solid symbols represent the no-shear case and open symbols are the sheared states for: $\Delta U=1$ (black) with $a\simeq 0.04$, $\Delta U=3$ with $a\simeq 0.10$ (blue), $\Delta U=5$ with $a\simeq 0.18$ (red).}
\label{figreol}
\end{figure}

We report in Fig. \ref{figreol}  (a) the reduced normal stresses $\theta_x\equiv P_{xx}/p, \theta_y\equiv P_{yy}/p$ measured in the simulations, as a function of the coefficient of restitution $\alpha$ and for several series with different shearing intensities. As we see, deviations from the NS behavior ($\theta_i\neq 1$) are small (less than $5\%$) even at moderately high values of shear (close to $a=0.2$). In this case, the coefficient of restitution seems not to have an important impact on the behavior of both  $\theta_x$ and $\theta_y$. In Fig. \ref{figreol} (b) we represent the value of the cross conductivity coefficient $\phi^*$. As we can see, the deviations from linear behavior (i.e., from $\phi^*(\alpha)=0$) are more significant here: up to a maximum of approximately $40 \%$ for the highest shear rate series represented here. However, as expected, the NS behavior is recovered for small shear rate (always less than $6.5 \%$ for the lowest non-null shear rate series). It is remarkable that, again, departures from NS behavior are stronger in the quasi-elastic limit than for more inelastic gases (i.e., more inelastic flows show less significantly non-Newtonian behavior here). This is consistent with the non-trivial observation in the previous section that $|\gamma|$ increases as we approach the quasi-elastic limit. As we already noticed, a higher $|\gamma|$ implies higher Knudsen number, and thus, our observations are self-consistent and not an artifact of the theory. The NS behavior ($\theta_i=1$, $\phi^*=0$) is strictly recovered for the reference states ($a=0$), for all values of inelasticity.


\section{Conclusion}

In summary, our results on the transport properties of our system strongly indicate that NS hydrodynamics may apply for steady flows of inert particles in a heterogeneous medium, as long as the relevant Knudsen number scales imposed by the boundary conditions in the problem do keep small. This is in no way different to the condition that is required to a molecular gas in order to obey hydrodynamics at NS order \cite{BCC95}. 

More generally, the fact that this result is independent of the degree of inelasticity supports the hypothesis that inelasticity by itself does not cause necessarily the breakdown of scale separation in granular gases such as it appears in the Homogeneous Cooling State. A recent work on the aging time to hydrodynamics for the homogeneous cooling state of the granular gas also supports this hypothesis \cite{VSK14}. The dynamics is actually a bit more complex and the energy source necessary to fluidize the system may cancel out the gradients inherently produced by inelasticity, bringing the granular gas back to the realm of NS hydrodynamics. In fact, theory-simulation comparisons seem to indicate that a part of the disagreement between theory and simulation can be ruled out when we consider an improve in the approximation to the reference distribution function in the Chapman-Enskog method (when a Sonine polynomial expansion to first order \cite{B35,T73} for the homogeneous cooling state \cite{NE98} is used instead of the Maxwellian \cite{GSM07}) and thus, the relatively small disagreement would not be due to an eventual failure of hydrodynamics but to the inherent weakness of the perturbative solution used in the Chapman-Enskog procedure \cite{CC70,BP04} for calculation of the transport coefficients.

Therefore, there is no a priori reason for concern respect to the validity of
hydrodynamics for granular gases. This would depend essentially on the specific
properties and geometry of the flow of interest, not just the inelasticity. 

Considering a granular gas fluidized by an active fluid should be regarded, not as an
artifact, but as a natural situation since, in the first place, rapid granular flows
are possible just because an energy input is able to compensate for inelastic cooling
(otherwise, clustering instabilities and total freezing of the dynamics occur
\cite{G03}). However, an interstitial fluid is a very frequent configuration for
granular dynamics problems present in nature \cite{B54}. Applications to biology
problems are therefore promising.  Moreover, it has been experimentally proved that
this type of fluctuating force like the one we use to model the active fluid reaches
local equilibrium with a granular particle \cite{OLDLD04}. Based on these
observations, our model for energy input from the active fluid should be closer, in
most cases, to experimental situations \cite{MSACVF13} where, interestingly, particle clustering instabilities also may occur.  

In addition, the results in the present work have helped us to produce further
results, since we have already applied NS hydrodynamics with a non-uniform stochastic
force also to problems and applications like granular segregation \cite{VGK14}, where
excellent agreement is found between granular impurity segregation criteria resulting
from NS theory and computer simulations. Finally, we plan to extend this work by
developing the corresponding non-Newtonian hydrodynamic theory and also by
calculating the linear stability criteria of theses base flows. On the other hand,
this work could be relevant for the study of transport in crowded active enviroments
\cite{bechinger16} where the time scales of the active fluids is faster than the
inert particles relaxation time. The existence of a well defined NS hydrodynamics in
heterogeneous active bath \cite{argun16}, is also an important result in order to
explain biological systems. Another possible future work is to study the effect of
sorting of active swimmers with inert particles in heterogeneous active media. Also
the study of the behaviour of particles,  mean square displacement, mobility,
diffusion coefficient depending on the media. A study of the hydrodynamics of more
complex forms of the volume force, for example with time dependency, such as for a
traveling wave \cite{sandor17}, is also a promising line of research.




\appendix


\subsection{Computer simulations}\label{comp}
As we said, we used for this work data from the direct simulation Monte Carlo method (DSMC) \cite{PS05} of the kinetic equation \eqref{rBE}. The standard DSMC algorithm consists of two basic steps: collisions (particle-particle collisions and, in this case, particle-boundaries collisions) and free drift \cite{B94}. The boundaries are modeled as hard walls, similarly to previous works. In the presence of a volume force, there is an additional step that accounts for the action of the fluctuating volume force. Implementation of this step, is described for instance in \cite{MS00,GSVP11}. However, for this work, we consider a more general situation where the noise intensity $\xi^2(y)$ is space-dependent. More specifically, the fluctuating force intensity has a profile obeying equation \eqref{profnoise}; i.e., of the form $\xi(y)^2\propto T(y)^{1/2}$. For the solution to be self-consistent, the temperature profile should be obtained, as we explained, from the differential equation \eqref{profT}, that is the same differential equation that the 'LTu' profiles obey \cite{VSG10}. For this reason, we have extracted simulation temperature profiles $T(y)_\mathrm{LTu}$ from LTu states obtained in a previous work \cite{VSG10}. We introduced them in condition \eqref{profnoise}, which defines a force intensity profile that is maintained constant in time during the simulation: $\xi(y)^2\propto T(y)_\mathrm{LTu}^{1/2}$. The initial temperature of the granular gas was however always set to the lower wall temperature (constant): $T_0(y)=T_-$. Once the simulation starts, and independently of the value of inelasticity, we always obtained after a transient a steady temperature profile $T(y)$ with the same form of a Fourier flow, as expected (see Fig. \ref{figT}). We have considered in all simulation figures presented in this work a system with $T_+/T_-=5$ and $\Delta L\equiv L=15\overline\lambda$ ($\overline\lambda=\sqrt{2}\pi\overline n\sigma^{d-1}$, $\overline n$ being the average density in the system). We also performed additional series with different $T_+/T_-=2, 10$ (not shown in figures), and checked that with this variation we obtained the same results.

\subsection{Navier-Stokes Transport coefficients}\label{coef}
The Navier-Stokes (NS) transport coefficients that we use in this work have been calculated by Garz\'o and Montanero in a previous work \cite{GM02}. Their expressions need to be inserted in Eqs. \eqref{ebal}, \eqref{profnoise} in order to complete the analytical solution of the hydrodynamic profiles, that result from the differential equations \eqref{Pij}, \eqref{difT} for sheared states and \eqref{profu}, \eqref{profT} for reference states (no shear). We write these transport coefficients in this appendix too, for the sake of completion. The reduced viscosity $\eta^*(\alpha)$, and reduced heat flux coefficients $\kappa^*(\alpha), \mu^*(\alpha)$ are \cite{GM02}

\begin{eqnarray}
& &\eta^*(\alpha)=\frac{1}{\nu^*_\eta}, \quad  \kappa^*(\alpha)=\frac{d-1}{d}\frac{1+c}{\nu^*_\kappa},\nonumber \\
& & \mu^*(\alpha)=\frac{d-1}{2d}\frac{c}{\nu^*_\mu}. \label{coefs}
\end{eqnarray}
 Here, $\nu^*_\eta$, $\nu^*_\kappa$, $\nu^*_\mu$ are the collisional frequencies associated to the transport coefficients and their expressions were calculated by Brey and Cubero, together with the expression for the reduced cooling rate $\zeta^*(\alpha)$ \cite{BC01}

\begin{eqnarray}
& &\zeta^*(\alpha)=\frac{d+2}{4d}(1-\alpha^2)\left(1+\frac{3}{32}c\right), \\
& &\nu^*_\eta=\frac{3}{4d}(1-\alpha+\frac{2}{3}d)(1+\alpha)\left(1-\frac{c}{64}\right), \\
& & \nu^*_\kappa=\nu^*_\mu=\frac{1+\alpha}{d}\left(\frac{d-1}{2}+\frac{3}{16}(d+8)(1-\alpha)\right) \nonumber \\
& & +\frac{1+\alpha}{d}\left(\frac{4+5d-3(4-d)\alpha}{1024}c\right). \label{freqs}
\end{eqnarray} The coefficient $c$ in \eqref{coefs}-\eqref{freqs} was calculated by van Noije and Ernst \cite{NE98} and has the expression

\begin{equation}
c=\frac{32(1-\alpha)(1-2\alpha^2)}{73+56d-3\alpha(35+8d)+30(1-\alpha)\alpha^2}.
\end{equation}

As we already mentioned, the additional (dashed) theoretical curves in Fig. \ref{figcoefs} correspond to the transport coefficients calculated with an improved approximation to the reference distribution function in the Chapman-Enskog method. In effect, in this improved procedure the reference distribution function an approximate form of the homogeneous state. This approximate form consists in an expansion in Sonine polynomials, where we only retain the first non-Maxwellian contribution. For more reference on this alternative procedure, please refer to \cite{GSM07} or \cite{GVM09}.

\acknowledgments{
Support of the Spanish Government through Grant FIS2010-12587 (F. V. R.), MTM2014-56948-C2-
2-P (A. L.)  is acknowledged. Partial support from FEDER funds and by Junta de Extremadura through Grant No. GRU10158 is also acknowledged (F. V. R.). A. L. thanks the hospitality of UEX where during a stay this work was done.
}

\bibliography{nLTu}

\end{document}